\documentclass[letter, mbf,wrapft]{ptptex}

\usepackage{wrapft}

\def\beqra{\begin{eqnarray}}
\def\eeqra{\end{eqnarray}}
\def\beqast{\begin{eqnarray*}}
\def\eeqast{\end{eqnarray*}}
\def\be{\begin{enumerate}}
\def\ee{\end{enumerate}}
\def\lag{\langle}
\def\rag{\rangle}

\def\beq{\begin{equation}}
\def\eeq{\end{equation}}

\notypesetlogo  

\title{
A Schematic Model of Leptons and Quarks\\
 Based on Majorana Partners
}

\author{
Kazuo {\sc Koike}
}
\inst{
Department of Natural Science,~Kagawa University,
Takamatsu 7608522, Japan
\\
}
\recdate{March 18, 2002
}
\abst{
 The possible extension of generation structure is investigated on the basis of
a composite model, in which leptons and quarks are composed of rishons $T$ with 
charge 1/3 and purely neutral $V$. 
In this sub-system, we regard $V$ as the origin of Majorana particles.
The seesaw mechanism implies the possible existence of heavy $V$, denoted as
$V_B$ in addition to ordinary $V$, denoted as $V_s$.
If we regard $V_B$ as a constitute block of leptons 
and quarks, a new generation scheme arises.
This new scheme satisfies the anomaly-free condition.
}
\begin{document}

\maketitle

\noindent
{\it 1. Introduction }

With the progress made in neutrino physics, the existence of Majorana particles now
seems almost certain.
It is well known that the extremely small mass of neutrinos is nicely explained by 
the seesaw mechanism,\cite{rf:GRSY} which is possible only for Majorana particles. 
Are there any Majorana particles in addition to neutrinos?
It should be noted that in a certain kind of composite model leptons and quarks 
contain purely electrically neutral constituents. In such a case, the 
additional particles appear in a generation structure, provided that their 
neutral components can be regarded as Majorana particles.
This note is concerning to this problem.
  We first give a short review of the rishon model,\cite{rf:rishon} which 
contains a purely 
electrically neutral constituent called a ``$V$ rishon". We regard $V$ as 
Majorana particle, and the seesaw mechanism concerning the $V$ rishon is formulated.
Then, a possible new structure of the generation is proposed, 
in which new heavy quarks and neutrinos appear. 

In this paper, we suppose the rishon is a purely mathematical entity or an 
entity whose dynamics are not yet known. In order to approach to its 
dynamics,
it is assumed that some frameworks of conventional field theory are partially 
applicable. 
It is also assumed that the gauge structure of the Standard Model appears 
as a result of the dynamics of a sub-level at the level of leptons and quarks, 
where GUTs and/or SUSY-like structure is partially realized.

\noindent
{\it  2. Model of leptons and quarks with neutral constituent }
 
It is known that only purely electrically neutral particles can be regarded 
as Majorana particles. Though the total charge of neutron is zero,  
it cannot be regarded as Majorana a particle, if ignoring its 
composite nature, because it has an intrinsic magnetic moment.
Thus, among observed leptons and quarks, only neutrinos can be Majorana particles. 
An additional possible Majorana particle appears in a 
certain kind of composite model of leptons and quarks. This is the $V$ rishon in 
the composite model of leptons and quarks proposed by Harari and Shupe.
\cite{rf:rishon} 
The rishon model is based on the elementarity of the electric charge, which seems 
to be absolutely conserved.
It should be noted that almost all quantum numbers 
concerning to conventional conservation laws are not conserved in GUTs
except the electric charge. Taking this point into account, the rishon model 
appears to be worthy of consideration.
Furthermore, the three generation structure in the ``cold-sector" is naturally 
explained by this model.\cite{rf:KK}
In the rishon model, all leptons and quarks
are three body systems of rishons $T$, with charge 1/3, and $V$, with charge 0, 
and color degree of freedom
of $u$ and $d$ quarks is realized by their configuration.
\cite{rf:rishon,rf:Geometrical}  
 The leptons and quarks in this model are given as\cite{rf:rishon} 
\beq
        { u =  TTV,~~
          d =  \bar{V}\bar{V}\bar{T},~~}
          {\nu}_e = {VVV,~~
          e^- =  \bar{T}\bar{T}\bar{T}}.
          \label{eq:Rishon}
\eeq
It should be noted that the $V$ rishon is purely electrically neutral, therefore
it is possible to introduce the Majorana interaction on $V$.

\noindent
{\it 3. The seesaw mechanism in the $V$ rishon as a Majorana particle}

For the $V$ rishon, let
us consider the Dirac-Majorana (D-M ) mass term 
\cite{rf:Pon}\tocite{rf:Bilen} in the simplest case of
one generation. 
We have\footnote{In a classification scheme based on the hyper-color and
color group
$ SU_3(H) \times SU_3(C) $,
the singlet property of the Majorana mass term may be destroyed if the rishon
belongs to the fundamental representation. The {\bf 8} representation 
seems to avoid this difficulty.
The details of the resolution of this problem will be discussed elsewhere.}

\begin{eqnarray}
{\cal{L}}^{\rm{D\mbox{-}M}}& = &
-\frac{1}{2} m_{L} \overline{(V_{L})^c} V_{L}
-m_{D}\bar{V}_{R}V_{L}
-\frac{1}{2} m_{R} \bar{V}_{R} (V_{R})^c~~+~~\rm{h.c.}
                  \nonumber\\
               & = &-\frac{1}{2}
\overline{{(V_{L})^c} \choose {V_{R}}}
{}~M~{{V_{L}} \choose {V_{R}^c}}
{}~~+~~\rm{h.c}.
\label{eq:DM}
\end{eqnarray}
\noindent Here
\beq
M = \left( \begin{array}{cc}
m_{L} & m_{D} \\
m_{D} & m_{R}
\end{array} \right),
\label{eq:MATm}
\eeq
\noindent where $m_{L}, m_{D} and m_{R} $ are parameters. For a symmetrical 
matrix $M$ we have
\beq
M=U~ m~ U^{\dag},
\label{eq:DIA}
\eeq
\noindent where $U^{\dag}U=1 $ and $m_{jk}=m_j \delta_{jk} $. From 
Eqs.~(\ref{eq:DM}) and
(\ref{eq:DIA}) we have
\beq
{\cal{L}}^{\rm{D-M}}=-\frac{1}{2} \sum_{\alpha=1}^{2}
 m_{{\alpha}}{\bar{\chi}}_{{\alpha}} \chi_{{\alpha}} ,
\label{eq:dia2}
\eeq
\noindent where
\begin{eqnarray}
V_{L}~~  =~~~{\cos{\theta}}\chi_{1L} & + & {\sin{\theta}}\chi_{2L},
\nonumber\\
(V_{R})^c  =  {-\sin{\theta}}\chi_{1L} & + & {\cos{\theta}}\chi_{2L}.
\label{eq:MIX}
\end{eqnarray}
\noindent Here $\chi_{1} $ and $\chi_{2} $ are fields of Majorana
 $V$ rishon with masses $ m_{s}~(a~~``small"~ mass)$, $m_{B}~(a~~``Big"~ mass) $, 
 respectively.
Assuming now that


\beq
m_{L}=0,~~~ m_{D}\simeq{ m_{F}},~~~m_{R}\gg{ m_{F}} ,
\label{eq:mass-sb1}
\eeq

\noindent where $m_{F} $ is a typical mass of the rishon,
particles with definite masses are distinguished as a very light Majorana
$V$ rishon with mass
$m_{s} \ll m_{F} $ and a very  heavy Majorana particle with mass
$m_{B}\simeq m_{R} $. The current $V$ rishon field, $V_{L} $, nearly
coincides with
 $\chi_{1L} $, and $ \chi_{2}\simeq {V_{R}~+~(V_{R})^c} $, 
because $\theta$ is extremely small.
It should be noted that $V_{L}$ and $(V_{R})^c$ represent the state corresponding
to that concerning the possible weak interaction, 
while  $\chi_{1} $ and $ \chi_{2}$ are the mass
eigenstates.
Hereafter, we use the notation
\beq
          V_s \equiv \chi_{1}~~~({\rm mass}~ m_s),~~~~~\\ 
          V_B \equiv \chi_{2}~~~({\rm mass}~ m_B) 
\label{eq:VsVB}
\eeq

In our scheme, the $V$ rishon $V_s$ is a Majorana
particle with masses much smaller than those of the other fermions.
The predictions for the $V$ rishon masses depend on the value of the
$m_{R}$ mass.

\noindent
{\it 4. New structure of the generation based on the Majorana partners 
}

The existence of $V_B$ in Eq.~(\ref{eq:VsVB}) implies the new 
generation structure with additional particles shown 
in Table~\ref{table:1}, where $TTV_s$, 
$\bar{T}\bar{V_s}\bar{V_s}$, $V_sV_sV_s$
and $\bar{T}\bar{T}\bar{T}$ represent the observed quarks and leptons,
$u$, $d$, $\nu$ and $e^-$, respectively.
In addition, new heavy quarks $TTV_B$, $~\bar{T}\bar{V_s}\bar{V_B}$ and
$\bar{T}\bar{V_B}\bar{V_B}$ and heavy neutrinos 
$V_sV_sV_B$, $V_sV_BV_B$ and$V_BV_BV_B$ appear. They are characterized by their
configurations containing the heavy $V$ rishon $V_B$.
The row corresponding to $\nu$ lists the mass eigenstates of neutrinos belonging 
to the first
generation, and $\nu_e$ the state concerning the weak interaction, 
which is realized as a superposition of mass eigenstates.
\begin{table}
\begin{center}
\caption{Configuration of leptons and quarks constituting a generation,
where the row corresponding to $\nu$ lists its mass eigenstates belonging to the 
first generation.
}
\label{table:1}
\begin{tabular}{cccccccc} \hline \hline
flavor   &standard      & $B$            & $BB$          & $BBB$ \\ \hline
$u$      &$TTV_s$       & $TTV_B$ \\
$d$      &$~\bar{T}\bar{V_s}\bar{V_s}$   & $~\bar{T}\bar{V_s}\bar{V_B}$
 & $\bar{T}\bar{V_B}\bar{V_B}$\\
$\nu$  &$~~V_sV_sV_s$   & $~~V_sV_sV_B$  & $~V_sV_BV_B$ & $~~V_BV_BV_B$\\
$~~e^-$    &$\bar{T}\bar{T}\bar{T}$\\ \hline   
\end{tabular}
\end{center}
\end{table}

It should be emphasized that the new scheme satisfies the anomaly-free condition
at the level of leptons and quarks;
\footnote{ In this paper, we have assumed that the gauge structure appears
at the level of leptons and quarks, as a result of the dynamics of the sub-system,
and the anomaly-free condition is formally realized at this level. In this sense,
the meaning of this condition in the sub-system seems to be rather ambiguous.}
that is, the condition
\beq
\sum{Q} = 0
\eeq
is satisfied.

It should be noted that
in the seesaw mechanism, if the mass of $V_B$ is of the order of the Planck
mass, the generation structure will not be affected in practice. The most 
interesting
case is in which the mass of $V_B$ is not extremely large,
for example, on the order of possible super-partners.
In such a case, heavy quarks and neutrinos appear at the level of leptons
and quarks.
In the special case in which the two Majorana particles have exactly the same
mass, $m_{V_s} = m_{V_B}$, they will behave as if a single Dirac particle did,
which is often called a ``pseudo Dirac particle".\cite{rf:PD} 

\noindent
{\it 5. Possible creation and decay modes of new particles}

Our model of leptons and quarks is an extension of the Standard Model that is based
on the possible Majorana property of the constituents of leptons and quarks.
As the fundamental dynamics of the sub-system are not yet known, we have assumed,
that the gauge structure of the Standard Model appears as a result the dynamics
of the sub-system,
at the level of leptons and quarks, and assumed that the quantum field theory 
is partially applicable in that sub-system.
Within this picture, the color structure of the Majorana partners 
$u_{B_1}$,~~$d_{B_1}$ ~and~ $d_{B_2}$~
is the same as that of ordinary quarks.
Thus, hadrons containing these particles will be created by pair creation such as
\beq
P + P \to {\rm hadron}( u_{B_1}\cdots) + {\rm hadron}
( \bar u_{B_1}{\cdot\cdot\cdot})\cdots,
\eeq
where $P$ and hadron$( u_{B_1}\cdots)$ represent the proton and hadron containing
the $u_{B_1}$ quark.

The decay mode of these Majorana partners depends depends strongly on the possible 
dynamics of the sub-system. If we naively assume that the electroweak doublet 
structure is realized 
in the sub-system 
$(T ~\bar V)$, sub-gauge bosons appear, and our Majorana partners are ``sterile" 
in the sense that they do not take part in the standard weak interaction.   
Thus, provided that this characteristic property remains unchanged in our system of 
leptons and quarks, the decay modes are restricted to pair annihilation 
and other interactions, such as rare processes, except to the standard weak 
interaction.
Therefore, it is expected that our Majorana partners have a considerably long life time.   

Further, it should be noted that model to explains
the proton-decay problem without difficulty. A typical decay mode is known as
\beq
P \to e^+  + {\pi}_0,
\eeq
which is caused due to the $X$ gauge boson in GUTs.
In our model, this process is realized through the same process as in GUTs,
\beq
u + u \to X \to e^+ + \bar d,
\eeq
where we have regarded the 6 rishon system as $X$.
It should be noted that the gauge bosons are assumed to be 6-body systems of rishons
in the first work on the rishon model.\cite{rf:rishon}
The suppression of this process will be reduced to that of the rearrangement of 
rishons or, equivalently, the largeness of the $X$ boson mass.

\noindent
{\it 6. Why are light quarks so light?}

It is well known that the light $u$ and $d$ quarks have very small masses compared 
with the typical electroweak mass scale. They are estimated to be on the order of
several MeV.\cite{rf:quark_mass}  
Through what mechanism is the smallness of these quark masses realized?

It should be noted that our scheme is consistent with the smallness of
the $u$ and $d$ quark masses, though it is not sufficient to explain smallness,
because we do not yet know the dynamics of these composite systems.
Furthermore, the seesaw mechanism in our scheme cannot explain the smallness 
of electron mass, in which no neutral constituent is contained. It should 
be remembered that the electron has the special
configuration $e^- =  \bar{T}\bar{T}\bar{T}$ in the rishon model.
In our opinion, a certain kind of statistics of order 3 may be concerned with
the realization of the small mass, together with the appearance of yet unknown 
dynamics.
Investigation of these points is a future problem.

\noindent
{\it 7. Discussion}

  In this paper, we have proposed a possible new scheme describing the generation 
realized as a result of the Majorana property of the $V$ rishon. 
It should be emphasized that our approach 
allows for the extension of 
the knowledge for existence-form of matter found in 
Majorana neutrinos\cite{rf:GRSY,rf:SK-Atmospher}
to other fundamental particles as a universal property. 
Properties of the fundamental leptons and quarks belonging to the first
generation are listed in Table~\ref{table:2}.

\begin{wraptable}{l}{\halftext}
\caption{Leptons and quarks belonging to the first generation.
}
\label{table:2}
\begin{center}
\begin{tabular}{cccccc} \hline \hline
&~$Q$        &flavor    & $B1$            & $B2$           & $B3$ \\ \hline
&$~2/3$    &$u$       & $u_{B_1}$ \\ 
&$-1/3$    &$d$       & $d_{B_1}$     & $d_{B_2}$ \\  
&$~0$      &~$\nu  $  & $~\nu_{B_1}$  & $\nu_{B_2}$  &$\nu_{B_3}$ \\
&$-1$     &~~$e^-$\\ \hline   
\end{tabular}
\end{center}
\end{wraptable}
It is noteworthy that the new heavy ``Majorana partners" 
$u_{B_1}$,~~$d_{B_1}$, ~~ $d_{B_2}$,~~$\nu_{B_1}$,~~$\nu_{B_2}$
and $\nu_{B_3}$ 
appear in our model.
Their possible existence should be made clear in the near future.  
Our work is based on the rishon model, which is still at a hypothetical level.
It should be emphasized, however, 
that this model is based on the simple concept of the elementarity of electric 
charge, and in this sense it is natural. In fact,
five types of electric charge appear in GUTs,\cite{rf:SU5} 
\beq
Q ~=~~0,~~~~1/3,~~~~2/3,~~~~1,~~~~4/3, 
\eeq
where the 4/3 charge is carried 
by $X$ gauge boson, which is 
considered to be the particle responsible for proton decay.

Then, as far as concerning to the charge units are concerned, it seems that 
GUTs are too complicated
to be the final form of the theory of elementary particles.

It should be noted that the standard formulation is not yet known in the 
rishon model, and, as mentioned above, it is probable that the rishon is beyond 
ordinary quantum field theory.\cite{rf:Maki} However, it seems to be meaningful to 
treat it in the framework of conventional field theory.
In this paper, we have assumed that  
the gauge structure of the Standard Model appears 
as a result of behavior of the level of leptons and quarks.
 It seems that there is another possible approach,
in which, the pre-electroweak structure 
appears in the rishon level, and a certain kind of confinement of 
pre-gauge bosons realizes the well-known electroweak structure at the level 
of leptons and quarks. In this case, the breakdown of symmetry
that causes the Majorana mass terms will occur successively after the 
breakdown of the electroweak structure of the sub-system.   
It should be emphasized that the seesaw mechanism itself does not
cause a small mass, but, instead, it is just a result
of breakdown of symmetry.
In order to understand the extremely small mass of neutrinos,
it may be necessary to assume further breakdown of symmetry
at the level of leptons and quarks that is responsible for the 
seesaw mechanism at this level.\cite{rf:induced_mass} 

Finally, is there really a sub-level below the level of leptons and quarks? 
In our opinion, such a sub-level surely exists, and disclosing it will make it 
possible
to predict theoretically quantities such as the Higgs coupling constants,
the magnitudes of symmetry breaking
$\lag\phi\rag$, the mass spectrum and mixing parameters of all particles, etc.
We restricted ourselves in this paper the  proposal of a possible 
new scheme for a generation with ``Majorana partners". The possible dynamical 
properties of rishon systems will be discussed elsewhere.
We conclude this paper by emphasizing that the possible existence of 
Majorana partners is probable, together with that of the super-partners in SUSY.
  


\begin{thebibliography}{99}

\bibitem{rf:GRSY}
M. Gell-Mann, P. Ramond and R. Slansky,
{\it Supergravity }, ed. Van Nieuwenhizen and D. Z. Freedman
(North Holland,1979).

T. Yanagida,
{\it Proc. of the Workshop on Unified Theory and Baryon Number
of the Universe}, Tsukuba, Ibaraki, Japan, 1979.

\bibitem{rf:rishon}
H. Harari,
Phys. Lett. {B \bf 86},(1979), 83.
   
M. Shupe, Phys.Lett. {B \bf 86},(1979), 87.
   
See also Y. Ne'emann,
Phys. Lett. {B \bf 82},(1979), 69.

\bibitem{rf:KK}

K. Koike,
\PTP{88,1992,81}.

\bibitem{rf:Geometrical}
For a  geometrical formulation based on the spin algebra formalism, see

E. Elbaz, 
Phys. Rev. {D \bf 34},(1986), 1612.

\bibitem{rf:Pon}
S. M. Bilenky and B. Pontecorvo,
Phys. Rep. {\bf 41} (1978), 225.

\bibitem{rf:Wol}
L. Wolfenstein, Phys. Rev. D {\bf 17} (1978), 2369.

S. P. Mikheyev and A. Yu. Smirnov,
{Yadern. Fiz.} {\bf 42} (1985), 1441.

\bibitem{rf:Bilen}
S. M. Bilenky and S. T. Petcov,
Rev. Mod. Phys. {\bf 59} (1987), 671.

\bibitem{rf:SU5}
The famous $SU$(5) GUT has the additional noteworthy property that it has
five fundamental charge units.

H. Georgi and S. L. Glashow,
\PRL{32,1974,438}.

\bibitem{rf:PD}
L. Wolfenstein,
Nucl. Phys.{\bf B186} (1981), 147.

{\bf 
As for the condition realizing the pseudo-Dirac particle, see

S. T. Petcov,
\PL{B 110, 1982,245}

C. N. Leung and S. T. Petcov,
\PL{B 125,1983,461}

S. T. Petcov and S. T. Toshev,
\PL{B 143,1984,175}

and  especially pages 690-694 in ref 7).
}

\bibitem{rf:quark_mass}
Particle Data Group, 
{\it Particle Physics Booklet.}

In this booklet, light quark masses are estimated as
$m_u$ = 1.5 - 5 MeV,~$m_d$ = 3 - 9 MeV.

For an early work in which the light quark mass is estimated as several MeV, see

Y.Tomozawa
Phys. Rev. B {\bf 186} (1969), 1504.

\bibitem{rf:SK-Atmospher}
Y. Fukuda et al.,
Phys. Rev. Lett. {\bf 85} (2000), 3999.

\bibitem{rf:Maki}
Z. Maki,
Prog. Theor. Phys. Suppl. No. 86 (1986), 313.

\bibitem{rf:Dynamical-R}
H. Harari and N. Seiberg,
\PL{98B,1981,269}.

\bibitem{rf:induced_mass}
A Majorana mass term may be induced by one or two loop diagrams. See

M. Doi, M. Kenmoku, T. Kotani, H. Nishiura and E. Takasugi,
\PTP{70,1983,1331}

\end{thebibliography}
\end{document}